%% file: main.tex
\documentclass[sigconf, nonacm]{acmart}

\input{datalabmacros}
\input{macros}

\begin{document}
\title{
ModelTables:  A Corpus of Tables about Models}

\author{Zhengyuan Dong}
\affiliation{%
  \institution{University of Waterloo}
  \city{Waterloo}
  \state{ON, Canada}
}
\email{zhengyuan.dong@uwaterloo.ca}

\author{Victor Zhong}
\affiliation{
  \institution{University of Waterloo}
  \city{Waterloo}
  \state{ON, Canada}
  }
\email{victor.zhong@uwaterloo.ca}

\author{Ren\'ee J. Miller}
\affiliation{%
  \institution{University Waterloo}
  \city{Waterloo}
  \state{ON, Canada}
  }
\email{rjmiller@uwaterloo.ca}

\input{src/abstract.tex}

\maketitle

\input{src/introduction}
\input{src/related}

\input{src/tableBenchmark}

\input{src/Statistics}
\input{src/experiment}
\input{src/applications}
\input{src/conclusion}

\section*{Acknolwedgements}
We acknowledge the support 
of the Canada Excellence Research Chairs (CERC) pro-
gram. Nous remercions le Chaires d’excellence en recherche
du Canada (CERC) de son soutien. %We thank the anonymous reviewers for their insightful comments and suggestions.

\bibliographystyle{ACM-Reference-Format}
\bibliography{main}

\end{document}

%% file: datalabmacros.tex
\usepackage{amsmath}
\usepackage{mathtools}
\usepackage{graphicx} % can better scale and rotate graphics than graphic package
\usepackage{url}
\usepackage{enumitem} % remove spaces inside 'itemize' [nolistsep]

\usepackage{stmaryrd} % for symbol \llceil
\usepackage{upgreek} % upright Greek letters; together with package "bm" -> upright bold letters
\usepackage{bbding} % for the \CheckmarkBold symbol

\usepackage{listings} % to get sql etc nicely formatted (source code listings)
\usepackage{tabularx} % table environment that allows to span a whole line
\usepackage{booktabs} % \toprule
\usepackage{multirow}
\usepackage{hhline} % cannot be overwritten by cellcolor
\usepackage{diagbox}
\usepackage{pgfplotstable} % powerful numerical table macros

\usepackage{xcolor,colortbl}    % !!! for some reason, overleaf needs xcolor separately, although that should already have been called by colortbl (10/2019)

\usepackage{tikz} % drawing figures
\usetikzlibrary{matrix}
\usetikzlibrary{calc}
\usetikzlibrary{math}
\usetikzlibrary{arrows.meta,positioning}
\usetikzlibrary{intersections, pgfplots.fillbetween}

\usepackage{easybmat} % \begin{BMAT} alternative syntax for math tabulars and also includes the command addpath which allows to draw a path between matrix elements.

\usepackage{adjustbox}
\def\eox{\unskip\kern 10pt{\unitlength1pt\linethickness{.4pt}$\diamondsuit${}}} % "\eox" command for end of example

\usepackage{rotating}		% for rotated text in tables, \begin{turn}{75}Accuracy\end{turn}

\usepackage{xspace} % for macros

\usepackage{subcaption} % recommended new package for subfigures (package subfig deprecated)
\newcommand{\hide}[1]{}

\usepackage[ruled,noend,linesnumbered]{algorithm2e} % boxruled, linesnumbered,ruled

\usepackage{setspace} % for setting line spacing

\DontPrintSemicolon     % does not print ";" at end of each line
\SetNlSty{}{}{}                % style of line numbers "\Setnlsty{small}{}{:}"
\SetAlgoInsideSkip{smallskip}   % distance between caption rule and beginning of algorithm
\SetAlFnt{\small}			% size of actual algorithm text\scriptsize\sf\rm
\SetAlCapFnt{\small}		% size of name of algorithm
\SetAlCapNameFnt{\small}
\usepackage{array}
\usepackage{makecell}
\usepackage[utf8]{inputenc}

\usepackage{aliascnt}  	
\usepackage{array}
\usepackage{makecell}
\usepackage[utf8]{inputenc}

\AtEndPreamble{%
    \hypersetup{colorlinks,
      linkcolor=purple,
      citecolor=blue,
      urlcolor=ACMDarkBlue,
      filecolor=ACMDarkBlue}}

\usepackage{tcolorbox}		% also allows to have shaded backgrounds around environments
\tcbuselibrary{breakable,skins}		% allow to break boxes, allow to use "enhanced jigsaw"

\tcbset{examplestyle/.style={
		enhanced jigsaw,	% if box is broken, don't show border at continuation
		colback=blue!08,	% !20	gray!10
		colframe=blue!08,	% !40
		arc=2mm,
		boxrule=0pt,		% 0pt
		left=1mm,
		right=1mm,
		left skip= 0mm,  % breaks ACM template unless specified for specific style (4/2019)
		right skip= 0mm, % breaks ACM template unless specified for specific style (4/2019)
		top=1mm,		% 0mm, -1mm
		bottom=1mm,		% problem in package with eq at end of env. just add: \tcbset{bottom=2mm} \tcbset{bottom=0mm}
		breakable,		% allows page break
		parbox = false,		% restores normal text behavior. ELSE: paragraphs are formatted slightly different as the main text. DOWNSIDE: unwanted side effects
		before={\par\pagebreak[0]\vspace{1mm}\parindent=0pt},		
		after={\par\pagebreak[0]\vspace{1mm}\parindent=0pt},				
		bottomrule = 0mm,
		boxsep = 0mm,					% common padding of hlengthi between the text content and the frame of the box, added to left, right, top, ...
		topsep at break=0pt,			% Additional vertical space at the top of middle and last parts in a break sequence
		bottomsep at break=0pt,			% Additional vertical space at the first and middle parts in a break sequence		
		pad at break=0mm,
		pad before break=1mm,		
		pad after break=1mm,		
		bottomrule at break=0mm,
		toprule at break=0mm,		
		}}

\usepackage{footnote}

\tcolorboxenvironment{example}{examplestyle}
\makeatletter
\DeclareRobustCommand*\uell{\mathpalette\@uell\relax}
\newcommand*\@uell[2]{
  \setbox0=\hbox{$#1\ell$}
  \setbox1=\hbox{\rotatebox{10}{$#1\ell$}}
  \dimen0=\wd0 \advance\dimen0 by -\wd1 \divide\dimen0 by 2
  \mathord{\lower 0.1ex \hbox{\kern\dimen0\unhbox1\kern\dimen0}}
}
\makeatother

\usepackage{marginnote}

\renewcommand{\epsilon}{\varepsilon} % nicer epsilon symbol
\usepackage{scalerel}
\usepackage{tikz}
\usetikzlibrary{svg.path}

\definecolor{orcidlogocol}{HTML}{A6CE39}
\tikzset{
  orcidlogo/.pic={
    \fill[orcidlogocol] svg{M256,128c0,70.7-57.3,128-128,128C57.3,256,0,198.7,0,128C0,57.3,57.3,0,128,0C198.7,0,256,57.3,256,128z};
    \fill[white] svg{M86.3,186.2H70.9V79.1h15.4v48.4V186.2z}
                 svg{M108.9,79.1h41.6c39.6,0,57,28.3,57,53.6c0,27.5-21.5,53.6-56.8,53.6h-41.8V79.1z M124.3,172.4h24.5c34.9,0,42.9-26.5,42.9-39.7c0-21.5-13.7-39.7-43.7-39.7h-23.7V172.4z}
                 svg{M88.7,56.8c0,5.5-4.5,10.1-10.1,10.1c-5.6,0-10.1-4.6-10.1-10.1c0-5.6,4.5-10.1,10.1-10.1C84.2,46.7,88.7,51.3,88.7,56.8z};
  }
}

\RequirePackage{etoolbox}
\DeclareRobustCommand\orcidicon[1]{\href{https://orcid.org/#1}{\mbox{\scalerel*{
\begin{tikzpicture}[yscale=-1,transform shape]
\pic{orcidlogo};
\end{tikzpicture}
}{|}}}}

%% file: macros.tex
\usepackage{array}
\usepackage{makecell}
\usepackage[utf8]{inputenc}

\newcommand{\arx}{arXiv}

%% file: src/abstract.tex
\begin{abstract}
We present \emph{ModelTables}, a benchmark of tables in Model Lakes that captures the structured semantics of performance and configuration tables often overlooked by text-only retrieval. The corpus is built from Hugging Face model cards, GitHub READMEs, and referenced papers, linking each table to its surrounding model and publication context. Compared with open data–lake tables, model tables are smaller yet exhibit denser inter-table relationships, reflecting tightly coupled model and benchmark evolution. The current release covers over 60K models and 90K tables. To evaluate model and table relatedness, we construct a multi-source ground truth using three complementary signals: (1) paper citation links, (2) explicit model-card links and inheritance, and (3) shared training datasets. We present one extensive empirical use case for the benchmark which is table search. We compare canonical Data Lake search operators (unionable, joinable, keyword) and Information Retrieval baselines (dense, sparse, hybrid retrieval) on this benchmark. Union-based semantic table retrieval attains 54.8\% P@1 overall (54.6\% on citation, 31.3\% on inheritance, 30.6\% on shared-dataset signals); table-based dense retrieval reaches 66.5\% P@1, and metadata-hybrid retrieval achieves 54.1\%. This evaluation indicates clear room for developing better table search methods. By releasing ModelTables and its creation protocol, we provide the first large-scale benchmark of structured data describing AI model.  Our use  case  of \emph{table discovery} in Model Lakes, provides intuition and evidence for developing more accurate semantic retrieval, structured comparison, and principled organization of structured model knowledge. Source code, data, and other artifacts have been made available at \url{https://github.com/RJMillerLab/ModelTables}.
\end{abstract}

%% file: src/introduction.tex
\section{Introduction}\label{sec:intro}

Many table corpora have been created and shared by researchers.  These include Webtables~\cite{cafarella2008webtables, eberius2015building, lehmberg2016large, korini2022sotab, peeters2024web}, 
Wikitables~\cite{DBLP:conf/kdd/BhagavatulaND13}, VizNet~\cite{DBLP:conf/chi/HuGHBZHK0SD19}, GitTables~\cite{hulsebos2021gittables}, and several Open Data corpora~\cite{DBLP:journals/pvldb/ZhuNPM16,nargesian2018table} containing tables crawled from Open Data Portals.  These table collections have been used to study a variety of data management tasks including semantic type annotation (determining the semantic type of an attribute), schema completion, table annotation, schema matching, entity matching, and table search (finding related tables).  Large corpora like GitTables are designed to be used in machine learning for table interpretation or understanding tasks~\cite{hulsebos2021gittables}.

In contrast, in this work, we present ModelTables, the first corpus of tables that describe AI models.  Unlike previous table corpora that have broad topical coverage and content, we focus on tables describing models.  To create our corpus, we make use of three types of resources.  First, a model repository (or model lake~\cite{DBLP:conf/edbt/PalBM25}) containing a large collection of models described by model cards~\cite{modelcardpaper, wolf2020transformers}  %\rjm{Zhengyuan:  please add citation to model card and data cards}  
and data cards which describe training data and benchmarks used in model development~\cite{gebru2021datasheets, lhoest2021datasets}.
Second, a code repository containing code related to the models.  Typically a model card will link to a code repository.    Third, a paper repository (or to be more specific for our benchmark, two paper repositories) containing scholarly papers that describe the models.  Model and data cards, along with code repositories can link to these papers.   Note that in our corpus, every table is associated with a model (or in the case of duplicates, a table may be associated with more than one model).

\subsection{From a Corpus to a Benchmark}

Notice that most of the  corpora mentioned are not benchmarks {\em per se}, in that they generally do not contain ground truth for any of the tasks they are designed to study.  Rather, researchers generally hand-curate subsets of these corpora or manually manipulate tables within them to create label benchmarks with ground truth.  

Consider, for example, the well-studied related table search.  Related tables are typically defined as tables that "include content that conceivably could have been in a single table"~\cite{DBLP:conf/sigmod/SarmaFGHLWXY12}, and with this idea in mind most work on related table search focuses on finding tables that can be unioned or joined~\cite{zhu2019josie,nargesian2018table,2023_dong_deepjoin,DBLP:conf/icde/BogatuFP020}.  
For benchmarking, a set of labeled tables is created (containing known related and non-related tables), for example the benchmarks for table union search TUS~\cite{nargesian2018table} and SANTOS~\cite{2023_khatiwada_santos} where labels are hand-curated or UGEN~\cite{pal2023generative} where labels are LLM generated.

\begin{figure*}[!htbp]
  \centering
  \includegraphics[width=\linewidth]
  {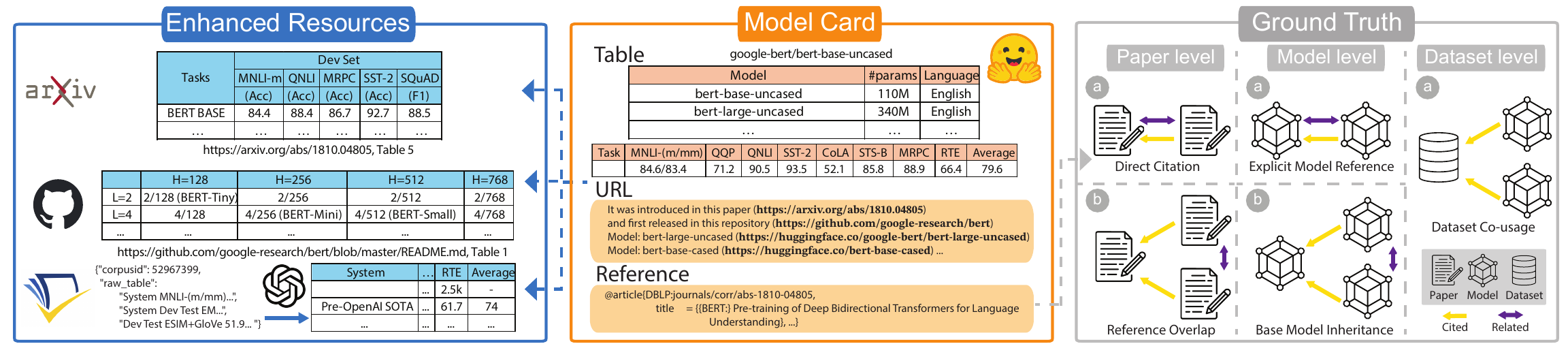}
  \caption{
The Model Lake Benchmark Setup Pipeline. This figure outlines our end-to-end, automated workflow for establishing the Model Lake benchmark. It illustrates the general-purpose pipeline for collecting, cleaning, and linking diverse tables, including the integration of a paper-level citation graph. Furthermore, it details how we construct a multi-level ground-truth for model relatedness by combining signals from paper citations, model lineage from model cards, and shared dataset metadata, providing a robust basis for evaluation.
}
  \label{fig:pipeline}
\end{figure*}

In contrast, we are publishing ModelTables with several possible baseline ground-truth annotations for table relatedness that are generated by model developers themselves.  And in contrast to other corpora, we propose using ModelTables to study a thematic notion of table relateness, rather than a structural one~\cite{DBLP:conf/sigmod/SarmaFGHLWXY12}.

Our {\em relatedness ground truth} comes from three sources.  First, model cards link to each other.  The first measure of relatedness is model card relatedness:  a model is related to any model that its model card links (via URL or a tag like base\_model).  
Second, data cards also link to each other and other models.
Third, paper relatedness is very well studied through the citations in papers and model cards link to papers.

\subsection{Benchmark Specifics}

We build a large-scale benchmark over the Hugging Face Model Lake\footnote{\url{huggingface.co}} and extract all tables from model cards or data cards.
Hugging Face is today, the largest open model repository %\rjm{is there any citation for this?}
~\cite{wolf2020transformers, lhoest2021datasets, modelcardpaper, gebru2021datasheets}.
Model (data) cards often refer to GitHub sites and we use GitHub as our code repository.  From GitHub README files (that are referenced by a model or data card), we also extract tables.  Finally, we extract tables from scholarly papers that are referenced in a model (or data) card or in a corresponding GitHub site.  In our work, we found many links to \arx\ papers\footnote{\url{arxiv.org}} along with bibtex entries and simple paper titles (often found in table captions).  For papers not in \arx\, we tried to find the paper source in semantic scholar.  We make use of the official Semantic Scholar Corpus\footnote{\url{https://www.semanticscholar.org/}}, a resource which is updated weekly and is accessible through Semanti Scholar API. Its metadata and citation extraction follow the S2ORC pipeline~\cite{lo-wang-2020-s2orc}, with commercial-scale refinements in the current deployed version.

\subsection{Contributions}\label{sec:contributions}
This work makes several contributions.
\textbf{(1)} We present the \emph{first large-scale benchmark of tables that describe AI Models}.  Unlike previous table benchmarks that are designed to have broad thematic content, our tables all describe models and therefore can be used in model understanding tasks.  
\textbf{(2)} We introduce a set of \emph{relatedness ground-truths} that are based on 
 verifiable relatedness signals from scholarly papers, model cards, and datasets, 
enabling our benchmark to be used for a variety of relatedness search tasks.
\textbf{(3)} We describe a reproducible pipeline for benchmark creation.  While our benchmark uses open models, code, and papers, our pipeline can also be used in-house to create a resource for private model lakes.  In such a scenario, the paper-relatedness would be replaced by project or team relatedness.\footnote{
LinkedIn Engineering Blog —
\href{https://engineering.linkedin.com/blog/2019/scaling-machine-learning-productivity-at-linkedin}{Scaling Machine Learning Productivity at LinkedIn};
Netflix TechBlog —
\href{https://netflixtechblog.com/supporting-diverse-ml-systems-at-netflix-2d2e6b6d205d}{Supporting Diverse ML Systems at Netflix.}
}
To date, there has been little work to help data-driven organizations manage and understand their own rich private model lakes.  Ours is a first step in demonstrating how structured data in a model lake can be used for model understanding.  
\textbf{(4)}  We analyze the benchmark and relatedness graphs and compare them to state-of-the-art table search corpora and benchmarks.
Our benchmark contains more related tables, with richer relatedness distributions that reflect the real nature of models (some models are hubs and related to many models and there is a skewed distribution with many models having few models they are related to).  
\textbf{(5)}  We empirically show the performance of existing related table search methods (from IR and data lakes) on model tables.  Our results indicate a research gap in finding thematically related tables. We show how our benchmark can be used to study new forms of table relatedness (beyond the typical structural or IR notions).
\textbf{(6)}  We discuss several use cases for the benchmark in model and table understanding.
Our results show that semantic search and table discovery uncover table relatedness from complementary perspectives, enabling richer understanding of model similarity.

\paragraph{Paper Organization}
In Section~\ref{sec:table-benchmark}, we describe our benchmark construction -- how we find and extract tables.  In Section~\ref{sec:ground_truth}, we describe our ground truth extraction.  We define three levels of ground truth all constructed from different signals provided by model developers.  In Section~\ref{sec:stats}, we analyze the benchmark and provide statistics on the tables, the different relatedness graphs and compare them to other table benchmarks (containing ground truth).  In Section~\ref{sec:experiments}, we consider related table search as an application for our benchmark.  We show that existing IR/DB table search approaches do not capture table relatedness well and point to new areas for research.  In Section~\ref{sec:apps}, we discuss other applications of our benchmark.  

%% file: src/related.tex
\section{Related Work}\label{sec:related}

\subsection{Data Discovery}

Retrieving structured tables in data lakes is typically posed as follows~\cite{DBLP:journals/vldb/ChapmanSKKIKG20,DBLP:journals/tweb/ZhangB21}: given a query table, the system must discover relevant tables in a large, heterogeneous repository based on schema compatibility, content overlap, or semantic similarity. Early work focused on joinable table discovery, where DeepJoin~\cite{2023_dong_deepjoin} uses pretrained language models to embed column values and perform nearest-neighbor search for both syntactic and semantic joins, while LSHEnsemble~\cite{DBLP:journals/pvldb/ZhuNPM16} and Josie~\cite{zhu2019josie} applies overlap-set indexing to efficiently identify tables sharing join keys. Unionable table search was addressed proposed by~\citet{nargesian2018table}.  SANTOS extends this work~\cite{2023_khatiwada_santos} using knowledge-base-driven semantic relations to improve union prediction and D3L by using regular expressions~\cite{DBLP:conf/icde/BogatuFP020}
In parallel, keyword-based table search methods index column names and metadata to support ad-hoc textual queries over the lake. 
Blend~\cite{esmailoghli2025blend} unifies and optimizes these classical data lake tasks.
Additionally, table relatedness techniques quantify semantic and structural affinities between tables by analyzing attribute co-occurrence statistics, thereby enhancing the precision of related-table retrieval~\cite{cafarella2008webtables}.

More recent research has introduced tasks that go beyond simple retrieval. Table reclamation ~\cite{fan2024gen} aims to recover missing rows or columns in a query table by selectively merging content from related tables, effectively “healing” incomplete datasets. Column disambiguation ~\cite{rastaghi2022robustnessentity} tackles inconsistent or ambiguous header names by linking them to canonical types or entities, improving interoperability across sources. Table augmentation ~\cite{yang2020generative} enriches an existing table with complementary attributes—such as geographic or demographic data—pulled from external repositories. By extending classical join, union, and keyword search with these new capabilities, data lakes become far more powerful and flexible for downstream analytics and AI workflows.

\subsection{Model Lake}

Model Lakes are still in an early, emerging phase~\cite{DBLP:conf/edbt/PalBM25,garouani2024model}.  \citet{DBLP:journals/corr/abs-2403-02327} present a vision for model lakes which includes  four core tasks: attribution, which traces a model’s outputs back to influential training data or internal features; versioning, which reconstructs directed model trees to reveal parent–child derivations; search, which locates models by intrinsic properties, output behavior, or metadata; and benchmarking, which evaluates models at scale using unified suites. 
\citet{garouani2024model} propose a three-zone architecture (ingestion, analysis, governance) with a rich metadata model for both data and models. 

Hugging Face is perhaps the largest open model lake currently.  Hugging Face supports keyword search over model names and full-text search over textual metadata.  Recently, they added a semantic search that uses LLM generated summaries of datasets and models and a similarity search over these summaries.\footnote{\url{https://huggingface.co/spaces/librarian-bots/huggingface-semantic-search}}
This is a black-box search and could not be included in our comparison.
Our work complements this by providing an open benchmark for evaluating their search and other semantic model and data set search approaches.  

\subsection{Scientific Tables}
In other fields, table-structured data has also been the subject of dedicated benchmarks and “lake” concepts. 
In the database community, GitTables ~\cite{hulsebos2023gittables} was created to support column-type detection and schema inference, providing a million-table corpus extracted from GitHub and a smaller labeled subset for evaluating semantic annotation models. 
In natural language processing, a variety of table-question-answering benchmarks have emerged.  
SciTab~\cite{lu2023scitab} focuses on fact-verification by pairing expert-validated scientific statements with their source tables, and TabLeX~\cite{desai2021tablex} evaluates both structure extraction and content recovery from table images—offering a broad spectrum of inference challenges. 
Meanwhile, in the scientific research domain, open-source data lakes such as SciSciNet~\cite{lin2023sciscinet} and CERN’s ESCAPE aggregate citation-related metadata and experimental results into unified repositories; although they adopt a “lake” architecture, their tasks center on large-scale scholarly analysis and data orchestration rather than fine-grained table understanding or model selection. 
More specifically, tables from scientific data lakes encompass diverse functional types—such as performance, configuration, and example tables—with corresponding tasks including semantic classification ~\cite{kim2012scientific, kruit2020tab2know},
structural classification~\cite{he2024towards}, and column-level annotation~\cite{hulsebos2021gittables} or classification~\cite{kruit2020tab2know, wu2023column}.

\subsection{Citation Graph}
Analysis of citation graphs is a well-studied area for tasks such as citation intent classification, link prediction, recommendation, and scientometric impact assessment. A wealth of platforms such as Crossref~\cite{hendricks2020crossref}, OpenAlex~\cite{priem2022openalex} and Semantic Scholar~\cite{fricke2018semantic} offer robust APIs and raw data dumps capturing billions of citation links across publications, funding grants, patents, clinical trials and media. The SciSciNet science-of-science data lake ~\cite{lin2023sciscinet} builds on these foundations by consolidating diverse sources into a unified analytical framework for studying research impact, collaboration patterns and the evolution of science. Early work centered on citation relationship classification exemplified by SciCite’s ~\cite{cohan2019structural} annotation of citation contexts as background, methodology, results or none and on citation link prediction using graph embeddings and neural network models. Recommendation systems exploit these link structures to surface relevant literature while citation based metrics quantify influence patterns and uncover emerging trends. Concurrently data citation research advanced by data citation~\cite{buneman2022data} treats datasets as first class nodes in the graph and enriches model selection signals in data lakes and model lakes by seamlessly integrating with versioning tasks. By unifying publication and dataset citations these approaches enable cross domain insights that bridge scholarly communication, data sharing and reproducibility.

%% file: src/tableBenchmark.tex
\section{Benchmark Construction}
\label{sec:table-benchmark}
This section details the construction of our model table benchmark, built upon a model lake. We begin by defining the foundational assumptions and scope that underpin the reliability and validity of the final benchmark. Following this framework, the practical construction proceeds through several key stages: comprehensive metadata extraction, rigorous table quality control, and the establishment of a ground truth for table relatedness, using signals from diverse sources like model cards, code repositories, and scholarly papers.
We describe our multi-stage table crawling pipeline (Figure~\ref{fig:pipeline}) designed to systematically gather tables and related metadata, including citations and URLs, enabling the formation of rich
relationships among models and datasets. 

\input{src/table_extraction}

\subsection{Extraction Quality Control}\label{sec:extractionQuality}
Our initial collection of tables, sourced from diverse platforms 
presents significant challenges in terms of quality and consistency. To create a high-quality and robust benchmark, our post-extraction pipeline refines the data through three key stages. We begin with \textbf{Quality Control} to correct low-level parsing patterns. This is followed by \textbf{Strategic Filtering} to remove entire tables unsuitable for analysis. Finally, the refined dataset undergoes \textbf{systematic Data Augmentation} to handle representational heterogeneity.

\paragraph{\bf Quality Control on Tables} 
Tables sourced from diverse platforms exhibit significant quality variance.   
These tables exhibit significant structural heterogeneity. Originating from formats like HTML and Markdown, their underlying structure is analogous to that of common web tables and CSV files and may exhibit  complex layouts with multi-level headers or merged cells. 
Our extract relies on widely-used SOTA table extraction methods~\cite{DBLP:journals/sigmod/HameedVPN25,christodoulakis2020pytheas} designed explicitly for processing highly heterogeneous tables.

Specifically, our pipeline handles several challenging edge cases. It resolves \emph{Column Alignment Issues} from merged or empty cells. It handles \emph{Internal Delimiter Conflicts} by specially parsing Markdown cells that contain the pipe character (`|`), a pattern that exists in some Hugging Face Model Cards.
\footnote{\url{https://huggingface.co/spacy/nb\_core\_news\_lg}} The pipeline preserves context from \emph{Cell-Level Annotations} by merging footnote text directly into its corresponding cell. It also stitches together fragmented \emph{Multi-Page Tables} and prunes \emph{Visual Formatting Artifacts} like empty rows.
Finally, for \emph{Special Characters and Symbols} (e.g., \texttt{\textbackslash lambda, \textbackslash  cdot}), our approach is to preserve them as-is. Given the vast number of domain-specific notations, preserving the original representation is more robust for maintaining semantic integrity than attempting a comprehensive normalization. These targeted corrections
ensure all tables are structurally sound.

\paragraph{\bf Strategic Filtering}
After correcting parsing errors, we apply a strategic filtering process to remove tables with significant structural issues or those unsuitable for column-wise analysis. Specifically, 
our LLM-based recovery over Semantic Scholar extracts yields tables of variable quality, making them appropriate only for an ablation study rather than our main results.  

\subsection{\bf Table Augmentation}
\label{sec:augmentation}
\begin{figure}[!htbp]
  \centering
  \includegraphics[width=\columnwidth]{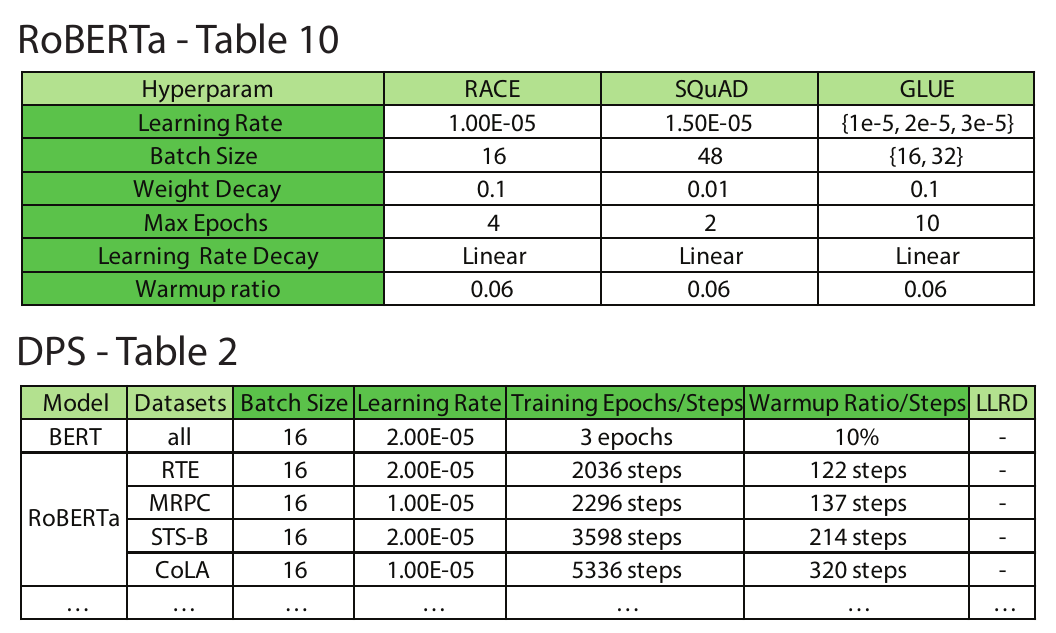}
  \caption{
  Tables from RoBERTa~\cite{liu2019roberta} and DPS~\cite{zhang2022fine} motivate augmentations: transposition for alignment, and header–cell fusion for semantic normalization% (e.g., “3 epochs”).
  } 
  \label{fig:permutation_example}
\end{figure}
A key challenge in model lakes is the significant heterogeneity in how tables are represented across sources. 
Scientific tables vary widely in orientation, formatting, and content encoding, which hinders straightforward matching and retrieval. 
To address this, we apply two  augmentation techniques designed to enhance table retrieval accuracy.  
Tables created by these augmentions can be optionally be added to the benchmark, depending on the use case.  

\paragraph{\bf Transpose Augmentation. } In model lakes, or more generally in scientific tables, authors may present information horizontally, for example, a table with a column for each of a set of benchmark datasets, where the column value is the performance of a configuration on that dataset.
Conversely, vertical layouts are frequently used when comparing different metrics for a single dataset, with metrics arranged as row labels.
These stylistic variations mean that tables representing similar information may have headers and key labels oriented differently. 
For instance, as shown in Figure~\ref{fig:permutation_example}, what appears as a column header in one table might be a row header in another. 
Given the tables from each of our four sources, we create a set of tables containing the transpose of each of these tables.

\paragraph{\bf Header-to-Cell Augmentation. }
In addition, we observe that some tables embed header semantics directly into the cell values. 
A typical case is columns that combine categorical labels with numerical data -- for example, a “Training Epochs/Steps” column may contain entries like “3 epochs” or “1500 steps,” merging quantity with unit in one cell. 
This practice, while enhancing human readability, introduces challenges for automated parsing and semantic alignment, as the same conceptual attribute might be split across different structural elements in different tables.
We can alleviate this and provide more semantic context for interpreting numeric values in tables but concatenating values with header names.
Given the set of tables from any of our four sources, we  create a header augmentation of this set where for each table every value in the table is concatenated with its attribute header name.

\input{src/modelRelatedness}

%% file: src/table_extraction.tex
\subsection{Table Extraction}
\label{sec:crawling_pipeline}
We construct our benchmark by collecting tables included in model card READMEs and additional tables from other resources referenced in the model cards.
We use the HuggingFace Model Card dataset\footnote{\url{https://huggingface.co/datasets/librarian-bots/model_cards_with_metadata} %Version 
250118} as our primary source,
HuggingFace is a large and popular model lake (with over 1M models at the time of our crawl).  It offers centralized hosting of model artifacts and rich metadata.
From each model card, we extract three core elements: embedded tables, URLs, and BibTeX references.  Then, for each URL and BibTeX entry, we detect its source platform (\arx\, GitHub, another Hugging Face model card) and apply tailored table scraping strategies.  The resulting benchmark contains 
table from four resources.
This extraction pipeline is formalized in Algorithm~\ref{alg:corpus_construction} and described below.

\input{src/tab_algo}

\paragraph{\bf Model Card Tables}
We first parse all tables directly embedded within model cards.  These tables are easy to parse and therefore very clean tables.  The tables are often configuration tables or performance tables, but we include all tables within the model cards.  

\paragraph{\bf GitHub Tables}
If a model card links to a GitHub repository, we download the README file and extract any tables found.  Like model card tables, these tables are easy to extract and also quite clean.  Some GitHub URLs point to large checkpoints or code repositories lacking README files or reliable places to find structured tables, and as a result do not contribute to our benchmark.

\paragraph{\bf \arx\ Tables}
If a model card links to one or more \arx\ paper (or has a BibTex file for an \arx\ paper), we would like to use the tables contained in these papers.  If the \arx\ link has an HTML version suitable for reliable table extraction, then we extract and use these tables from the markup.  At this time, we found that table extraction directly from PDFs did not yield high quality tables, so our benchmark does not include these.  
To avoid including irrelevant papers and tables,
we explicitly exclude the placeholder paper link\footnote{\url{https://arxiv.org/abs/1910.09700}}, found in the Hugging Face's template model card that many people use as a starting point and forget to delete\footnote{\url{https://huggingface.co/templates/model-card-example}}.

\paragraph{\bf Semantic Scholar Tables}
If a model card links to a paper in Semantic Scholar\footnote{\url{https://www.semanticscholar.org}}, 
we leverage the Semantic Scholar Open Research Corpus (S2ORC) ~\cite{lo-wang-2020-s2orc}. S2ORC processes PDF and LaTeX files through its pipeline and saves identified tables, figures, and text sections in its public data dump. However, these extracted tables are unstructured raw text, lacking a row-column structure. Therefore, after obtaining this raw table text from the S2ORC data, we pipe it into a Large Language Model (LLM) for parsing and refinement. The LLM's task is to convert these unstructured text blocks into a clean, uniformly structured table format.

%% file: src/tab_algo.tex
\begin{algorithm}[t]
\caption{Table Corpus Construction from Model Lake}
\label{alg:corpus_construction}
\KwIn{Set of all model cards $\mathcal{MC}_{all}$ from Hugging Face}
\KwOut{Curated set of tables $\mathcal{T}$, linked to its source $\mathcal{MC}$}

$\mathcal{T} \gets \emptyset$\;
$\mathcal{MC}_{valid} \gets$ Filter $\mathcal{MC}_{all}$ for cards with paper and tables\;

\ForEach{$mc \in \mathcal{MC}_{valid}$}{
    $T_{card} \gets \text{ExtractTables}(mc.\text{readme})$\;
    Add all tables in $T_{card}$ to $\mathcal{T}$, linking them to $mc$\;

    $P_{refs} \gets \text{ExtractPaperReferences}(mc.\text{readme})$\;
    \ForEach{$p \in P_{refs}$}{
        $T_{arxiv} \gets \text{ExtractTables}(\text{GetArxivContent}(p))$\;
        Add all tables in $T_{arxiv}$ to $\mathcal{T}$, linking them to $mc$\;

        $T_{s2orc\_raw} \gets \text{GetS2ORCTextTables}(p)$\;
        $T_{s2orc\_proc} \gets \text{RecoverTableStructure}(T_{s2orc\_raw})$\;
        Add all tables in $T_{s2orc\_proc}$ to $\mathcal{T}$, linking them to $mc$\;
    }

    $G_{urls} \gets \text{ExtractGitHubURLs}(mc)$\;
    \ForEach{$g \in G_{urls}$}{
        $T_{github} \gets \text{ExtractTables}(\text{GetReadme}(g))$\;
        Add all tables in $T_{github}$ to $\mathcal{T}$, linking them to $mc$\;
    }
}
\Return{$\mathcal{T}$}
\end{algorithm}

%% file: src/modelRelatedness.tex
\section{Model and Table Relatedness}
\label{sec:ground_truth}

In addition to extracting tables, we also extract model ids and paper ids.  Hence, our corpus contains a heterogeneous graph containing tables, models and papers as nodes.   All tables are associated (conceptually via an edge) with at least one model (more if there are duplicate tables).  A model may be associated with one or more papers (through its model card, associated data card, or associated code repo).  And every paper is associated with at least one model.  These direct "extraction" links or edges provide a way of relating tables and models.  In this section, we explore additional semantic for relatedness.  
Specifically, from this heterogeneous graph, we wish to exploit both the direct associations and possibly indirect associations to build several relatedness graphs that may be used in table and model search and understanding tasks.  

Our central hypthesis is that ModelTables can be used for a variety of table and model understanding tables.  For example, by finding and integrating tables related to a model, we can better answer complex data science questions such as "What is the best model for my task?"  Hence,
a key challenge is to find related tables and related models within our benchmark.   
This raises a fundamental question: how do we determine if two models are truly related? To answer this, we construct a few possible model relatedness graphes.
The core concepts guiding the construction of these graphs
are formally defined in Table~\ref{tab:concept}.

\begin{table}[h!]
\centering
\caption{Definitions for Model and Table Relatedness.}
\label{tab:concept}
\begin{tabular}{p{\dimexpr\linewidth-2\tabcolsep}}
\toprule
\textbf{Concept, Definition, and Detection} \\
\midrule
\textbf{Model Relatedness:} A connection between models based on their model card or training data. It is modeled via signals at the paper level (citations, shared references), model level (direct model card links or lineage indicated by model tags), and dataset level (common datasets). \\
\addlinespace
\textbf{Table Relatedness:} An inferred relationship between tables. %used as a proxy for model relatedness. 
It is modeled using the location of extraction and by inheriting the relatedness status from the source models or papers associated with each table. \\
\bottomrule
\end{tabular}
\end{table}
\label{sec:ground_truth}

We assess model relatedness across three key levels, as illustrated in %Figure~\ref{fig:relation_levels}. 
Figure~\ref{fig:pipeline}, paper, model, and dataset.

\subsection{Paper Relatedness}\label{sec:paper-related}
First, at the paper level, we obtain each paper’s references and citations from Semantic Scholar data. 
Based on this information, we identify whether two papers directly cite each other or share overlapping reference lists to determine their relatedness.

More formally, two models $m_i$ and $m_j$ are paper related ($R_{\text{paper}}$) if any of their associated papers are related through:
    \begin{enumerate}[label=(\alph*)]
        \item \text{Direct}:  \textit{Direct Citation} — one paper directly cites the other; %or 
        \item \text{Overlap}:  \textit{Reference Overlap} — their reference lists share at least one common cited work.  Reference overlap is computed over reference lists to capture both direct and latent citation relationships.  As an example two papers published almost concurrently may be very related but may not cite each other.  However, reference overlap will reveal the relationship.  
    \end{enumerate}

Citation metadata are retrieved from the Semantic Scholar API\footnote{\url{https://www.semanticscholar.org/product/api/tutorial}} 
and include two key attributes: citation \textit{intent} and \textit{isInfluential}~\cite{cohan2019structural,valenzuela2015identifying}. 
The \textit{intent} citations come from a \textit{Methodology} or \textit{Results} sections of papers and are considered more meaningful than (for example) citations contained in a background or related work section, which is derived from a multi-class classification model that categorizes the rhetorical role of each citation.
The \textit{isInfluential} is a binary indicator to detect if the citation  appears to be important, produced by a binary classification model that integrates multiple textual and citation-level factors, such as citation frequency, section placement, and contextual emphasis~\cite{cohan2019structural,valenzuela2015identifying}.

Of the various types of citation relationships, direct citations are the stricter is some sense than overlap citations, but these two types of relationships will form different relatedness graphs between models (or their tables).  Within each set (direct vs. overlap) citations, we can restrict our consideration to citations that have both the {\it isInfluential} and {\it Intent} attributes, have one of these, or consider any citation.  
Hence, the paper-related edges of our model graph can be used to form eight possible relatedness graphs based on the citation type:  {\tt Direct (Intent $\cap$ Influential); Direct (Intent); Direct (Influential); Direct;  Overlap (Intent $\cap$ Influential); Overlap (Intent); Overlap (Influential); Overlap}.

When using paper relatedness ($R_{\text{paper}}$) we refer to the use of all (\text{Direct} $\cup$ \text{Overlap}) edges, unless otherwise indicated.  In practice, we expect a use case will only use a subset of these.  

\subsection{Model Card Relatedness}\label{sec:model-related}

Second, at the model card level, we consider the base model tags provided in model cards to determine whether one model is an ancestor of another, as well as whether two models share the same ancestor.  We will consider such models related.
Additionally, if one model card contains a direct link to another model card, these two models are regarded as related.

More formally, two models are model related ($R_{\text{model}}$) if either:
    \begin{enumerate}[label=(\alph*)]
        \item \textit{Explicit Model Reference} — one model card explicitly links to another model card; or
        \item \textit{Base Model Inheritance} — a model declares another as its \texttt{base\_model} (or equivalent) in the metadata, 
        indicating lineage through fine-tuning, adapter injection, quantization, or model merging; or
        \item \textit{Shared Ancestor} - two models share the same \texttt{base\_model}.
    \end{enumerate}
Hyperlinks are resolved to canonical identifiers following the pattern \texttt{https://huggingface.co/\{org\_id\}/\{repo\_id\}}, 
and shorthand references (e.g., \texttt{gpt2} instead of \texttt{openai-community/gpt2}) are normalized using download frequency statistics.

\subsection{Data Card Relatedness}\label{sec:data-related}

Finally, at the dataset level, two models are considered related if they are trained on or can be inferred to operate on the same dataset. 
More formally, two models are dataset related $R_{\text{dataset}}$ if their model cards reference at least one common dataset (via dataset tags or URLs). 
    Dataset identifiers are extracted from model card READMEs (e.g., 
    \texttt{https://huggingface.co/datasets/\{org\_id\}/\{repo\_id\}}) 
    and validated against the official Hugging Face dataset metadata\footnote{\url{https://huggingface.co/datasets/librarian-bots/dataset_cards_with_metadata}, version~
    250428} 
    to ensure canonical alignment.
    
\subsection{Model and Table Relatedness Graphs}

Each table $t_i$ inherits the relatedness of its associated model(s) $M(t_i)$; hence, table-level relationships are defined as:
\begin{equation}
\begin{split}
R_T(t_i, t_j) =\;& R_{\text{paper}}(M(t_i), M(t_j)) \\
 &\cup R_{\text{model}}(M(t_i), M(t_j)) 
 \cup R_{\text{dataset}}(M(t_i), M(t_j))
\end{split}
\end{equation}
However, the benchmark is constructed so that the different types of relatedness may be used in concert or individually.  
We then operationalize these definitions when constructing the ground-truth relatedness matrices (the various $R$), as detailed in Algorithm~\ref{alg:gt_inference}.

\input{src/gt_algo}

\label{sec:aug_citation}

%% file: src/gt_algo.tex
\begin{algorithm}[t]
\caption{Table Ground Truth Inference}
\label{alg:gt_inference}
\KwIn{Tables $\mathcal{T}$; Citation graph $G_{cite}$; Filter config $F_{config}$}
\KwOut{Binary relatedness matrices $\{R_{\mathcal{T}}^{paper}, R_{\mathcal{T}}^{model}, R_{\mathcal{T}}^{dataset}\}$}

\SetKwFunction{FPaperRel}{ArePapersRelated}
\SetKwFunction{FModelRel}{AreModelsRelated}
\SetKwFunction{FDatasetRel}{AreDatasetsRelated}
\SetKwProg{Fn}{Function}{:}{}
\SetKwProg{Pn}{Procedure}{:}{}

\Fn{\FPaperRel{$m_a, m_b, G_{cite}, F_{config}$}}{
    \ForEach{$(p_a, p_b) \in \text{GetPapers}(m_a) \times \text{GetPapers}(m_b)$}{
        $(R_a, R_b) \gets \text{GetRefs}(p_a, p_b, G_{cite})$\;
        $(R_a, R_b) \gets \text{FilterByIntent}(R_a, R_b, F_{config}\text{.intent})$\;
        \If{$F_{config}\text{.influence}$}{
            $(R_a, R_b) \gets \text{FilterByInfluence}(R_a, R_b)$\;
        }
        \uIf{$F_{config}\text{.type} = \text{'Direct'}$}{
            \If{($p_b \in R_a$) \textbf{or} ($p_a \in R_b$)}{\Return{true}}
        }
        \uElseIf{$F_{config}\text{.type} = \text{'Overlap'}$}{
            \If{HasOverlap($R_a, R_b$)}{\Return{true}}
        }
    }
    \Return{false}\;
}

\Fn{\FModelRel{$m_a, m_b$}}{
    \Return{HasLineage($m_a, m_b$) \textbf{or} HasSharedBaseModel($m_a, m_b$)}
}

\Fn{\FDatasetRel{$m_a, m_b$}}{
    \Return{HasSharedDataset($m_a, m_b$)}
}

\vspace{0.3em}
\Pn{Main}{
    Initialize $R_{\mathcal{T}}^{paper}, R_{\mathcal{T}}^{model}, R_{\mathcal{T}}^{dataset}$ as zero matrices\;
    \ForEach{$(t_a, t_b) \in \mathcal{T} \times \mathcal{T}$}{
        \ForEach{$(m_a, m_b) \in \text{GetModels}(t_a) \times \text{GetModels}(t_b)$}{
            \If{%$R_{\mathcal{T}}^{paper}(t_a, t_b) = 0$ \textbf{and} 
            \FPaperRel{$m_a, m_b, G_{cite}, F_{config}$}}{
                $R_{\mathcal{T}}^{paper}(t_a, t_b) \gets 1$\;
            }
            \If{%$R_{\mathcal{T}}^{model}(t_a, t_b) = 0$ \textbf{and}
            \FModelRel{$m_a, m_b$}}{
                $R_{\mathcal{T}}^{model}(t_a, t_b) \gets 1$\;
            }
            \If{%$R_{\mathcal{T}}^{dataset}(t_a, t_b) = 0$ \textbf{and} 
            \FDatasetRel{$m_a, m_b$}}{
                $R_{\mathcal{T}}^{dataset}(t_a, t_b) \gets 1$\;
            }
        }
    }
    \Return{$\{R_{\mathcal{T}}^{paper}, R_{\mathcal{T}}^{model}, R_{\mathcal{T}}^{dataset}\}$}\;
}
\end{algorithm}

%% file: src/Statistics.tex
\section{Benchmark Statistics}\label{sec:stats}
In this section, 
we analyze the characteristics of the benchmark.

\subsection{Table Statistics}
\begin{figure}[!htbp]
\centering    \includegraphics[width=\linewidth]{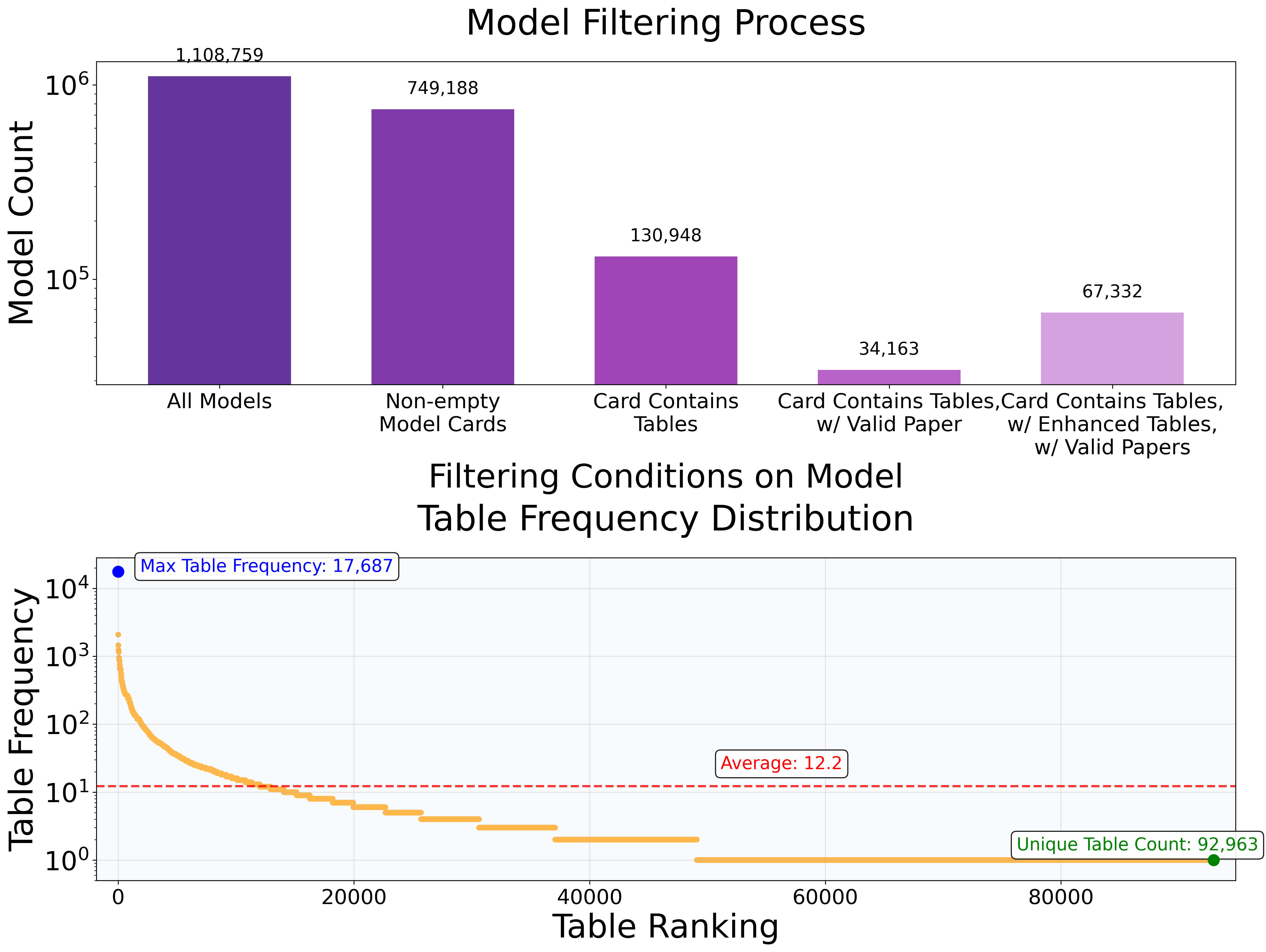}
\caption{Model- and table-level filtering of the model lake.
Top: Stepwise filtering of models by card completeness, table presence, and valid references, yielding the final benchmark subset. Bottom: Table frequency distribution showing a long-tailed reuse pattern that motivates deduplication.
}
\label{fig:model-level-stats}
\end{figure}
\begin{figure}[!htbp]
\centering    \includegraphics[width=\linewidth]{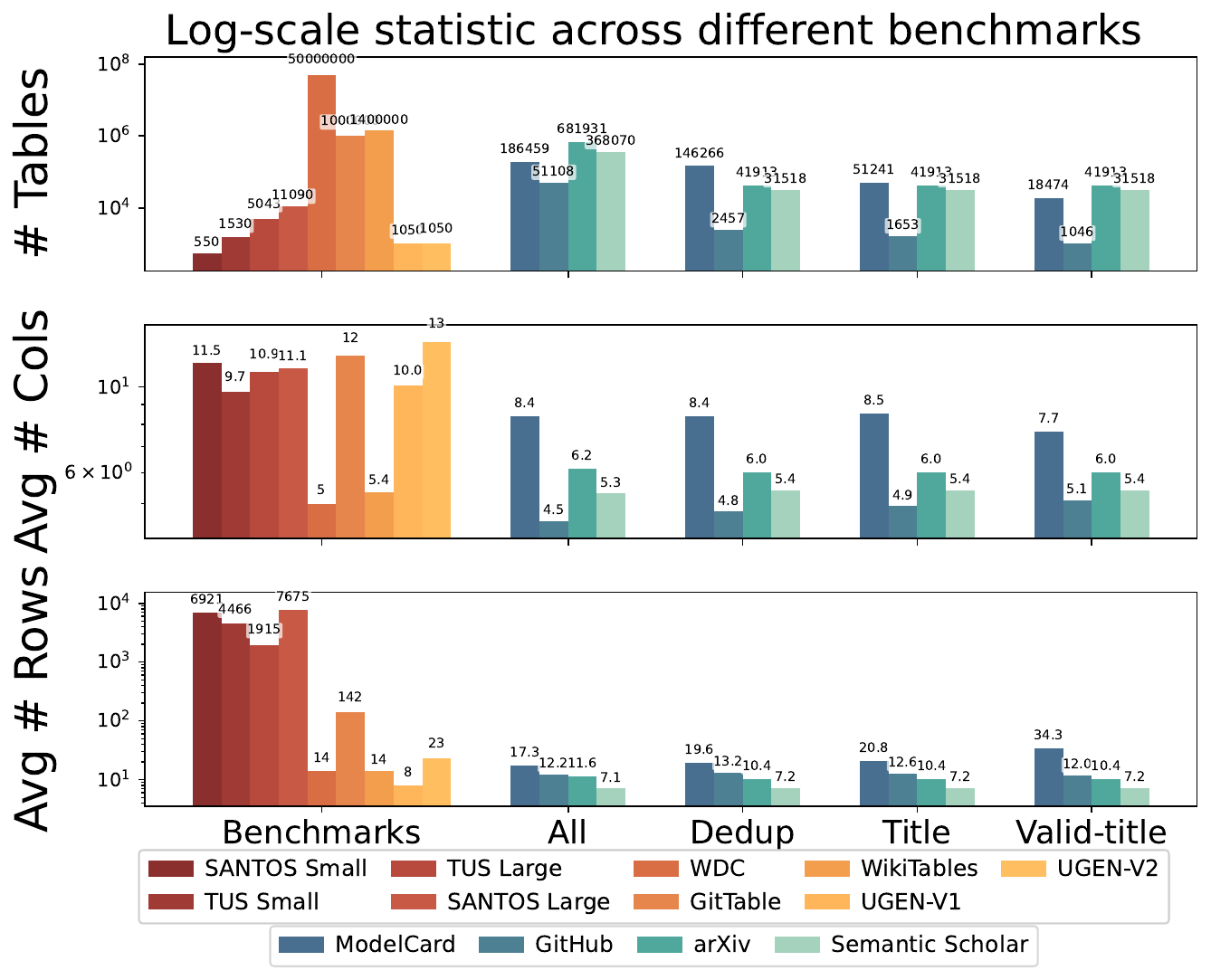}
\caption{
Our benchmark contains a larger quantity of smaller tables compared to %existing 
data lakes, a structural profile driven by the distinct characteristics of its academic sources.
}
\label{fig:stats}
\end{figure}

\paragraph{\bf Model-Level Quality Filtering. }
We began our benchmark construction from a model lake collection of over 1.1M models and their associated model cards. As shown in Figure~\ref{fig:model-level-stats}(top), about 68\% of these model cards are non-empty, containing at least minimal documentation. Among them, roughly 17\% include extractable tables that summarize model characteristics or results.
To create a high quality corpus of models with scholarly publications, we further restrict to model cards that reference valid research papers. This step substantially narrows the pool, as many community-submitted cards lack formal paper links. 
To mitigate this sparsity, we integrate other resources, including GitHub READMEs, arXiv, and Semantic Scholar, to recover missing table or citation information. After this augmentation, around 60K models satisfy all quality criteria and form the core of our benchmark.

\paragraph{\bf Table-Level Duplication Analysis. }
The bottom of Figure~\ref{fig:model-level-stats} analyzes the resulting table corpus by showing the frequency of table occurrences after extraction. We observe a pronounced long-tailed distribution: most tables appear only once, while a few popular or widely reused ones recur many times across model cards. This aligns with our intuition that valid model cards typically contain one or several unique tables, while occasional anomalies correspond to reused or mis-parsed tables. Such evidence validates our deduplication strategy, ensuring that the benchmark retains representative yet non-redundant tables.

\paragraph{\bf Benchmark Filtering Overview. }
Our goal is to refine the raw corpus of extracted tables into a high-quality benchmark suitable for computing table- and model-level relatedness. Figure \ref{fig:stats} (blue/green shades on the right) summarizes how the dataset contents changed with respect to some table filters.

\paragraph{\bf Stage A - All -> Dedup (Deduplication insights).}
Removing redundant tables eliminates the long tail of boilerplate or template-derived tables and retains roughly 17\% of the original corpus. Average column counts remain nearly unchanged, while the average number of rows increases slightly, evidence that duplicates are predominantly small, repetitive tables rather than structurally rich ones. Duplication intensity differs strongly by source:  the code resource GitHub shows heavy reuse, whereas our two paper sources like \arx\ and Semantic Scholar exhibit much less redundancy. 

\paragraph{\bf Stage B - Dedup -> Title (Contain references).}
The next step links each table to at least one textual anchor, such as a model-card title, a GitHub README heading, or a paper title. This constraint mainly removes code repository tables lacking citations or bibliographic context while leaving most paper tables intact. Conceptually, the corpus transitions from raw structural artifacts to citation-anchored tables, enabling downstream computation of paper- and model-level relationships.

\paragraph{\bf Stage C - Title -> Valid-Title (Scholarly validation).}
Titles are then validated against an external scholarly index (Semantic Scholar) to confirm that each refers to a verifiable research artifact. After this filtering, the benchmark stabilizes at roughly $10^5$ tables. Within this subset, Model-Card tables that pass validation become noticeably larger (more rows), but  narrower (fewer attributes) as larger performance tables predominate.
%—longer performance lists with concise schemas—while wide, template-style cards are filtered out. 
Roughly two-thirds of the model card titles and about one-third of GitHub references fail this match, whereas nearly all papers titles resolve cleanly as expected. The surviving corpus thus reflects academically grounded, structurally consistent content suited for quantitative relatedness analysis.

\paragraph{\bf Comparison to existing Corpora and Benchmarks}
Figure \ref{fig:stats} (orange/red shades on the left) contains several popular table corpora and a few data lake benchmarks (the later containing ground truth).  
Our benchmark exhibits a distinct structural and semantic profile compared to classic data-lake benchmarks such as SANTOS~\cite{2023_khatiwada_santos} and TUS~\cite{nargesian2018table} and to newer LLM generated benchmarks UGEN V1/2~\cite{pal2023generative}. We obtain many more, but smaller tables (comparable in size to the smaller UGEN tables).
While the overall statistics differ, the two types of benchmarks remain complementary.  Recall the data lake benchmarks are either hand-curated from open data so that ground-truth relatedness is know (TUS/SANTOS) or LMM generated (UGEN).
As we report below these benchmarks differ more in the characteristics of the ground truths (relatedness distributions), see Section~\ref{sec:ground-truth-statistics}.
Traditional data lakes capture large-scale structural diversity, whereas our benchmark emphasizes academically structured precision. Together, they offer a broader landscape for exploring table discovery, integration, and evaluation methods across heterogeneous domains.
In comparison, the (unlabeled) larger Web corpora (WDC, WikiTables, GitTables) contain many more tables.  Excluding GitTables, these are typically smaller tables as these come from web pages, and mass-collaboration pages.

\subsection{Table Relatedness Statistics}
\label{sec:ground-truth-statistics}
We have already examined the filtering quality of models and tables (Figure~\ref{fig:model-level-stats}, ~\ref{fig:stats}), ensuring that each retained table is linked to a valid paper.
In this section, we turn to the ground-truth relatedness statistics derived from our three types of relatedness.

\paragraph{\bf Paper Relatedness Ground Truth Density.}
We begin by quantifying the density of the paper relatedness ($R_{\text{paper}}$) of tables for the eight different subsets of $R_{\text{paper}}$.
Table \ref{tab:citation-density-summary}.
The raw overlap graph connects over 40\% of all table pairs, reflecting broad, but noisy linkage.
Applying influence or intent filters reduces this density to 10–25\%, and combining both yields a compact around 8\% high-confidence core.
In contrast, direct citation graphs remain substantially sparser (around 3–8\%), indicating higher precision, but narrower coverage.
These statistics confirm that our ground-truth graph preserves sufficient connectivity for meaningful discovery tasks without collapsing into an unrealistically dense structure.
The contrasting densities highlight two complementary capabilities of the benchmark.  
Overlap-based graphs recover indirect semantic relationship. While direct graphs represent explicit citation lineage.
Together they form a precision–recall trade-off: overlap offers breadth and higher recall; direct citations offer precision and reliability.

\begin{table}[!htbp]
  \centering
  \caption{Paper Relatedness Density across Citation Types. Applying influence and intent filters sparsifies the graph, %isolating higher-precision related paper subsets, 
  while overlap-based methods capture indirect relationships resulting in a much denser graph.} 
  \label{tab:citation-density-summary} 
  \begin{tabular}{lcc} 
\toprule 
\textbf{Citation Type}                & \textbf{Nonzero Edges}       & \textbf{Density} \\
\midrule 
Overlap                 & $3.71\times10^9$     & $43.03\%$ \\
Overlap (Intent)        & $2.23\times10^9$     & $25.83\%$ \\
Overlap (Influential)   & $8.59\times10^8$     & $9.94\%$ \\
Overlap (Intent $\cap$ Influential) 
&$6.95\times10^8$     & $8.04\%$ \\
Direct                       & $7.06\times10^8$     & $8.17\%$ \\
Direct (Intent)              & $4.98\times10^8$     & $5.76\%$ \\
Direct (Influential)         & $3.01\times10^8$     & $3.48\%$ \\
Direct (Intent $\cap$ Influential) & $2.67\times10^8$     & $3.09\%$ \\
\midrule 
Model & $1.71 \times 10^7$ & $0.19\%$ \\
Dataset & $3.51 \times 10^7$ & $0.41\%$ \\
All (Direct $\cup$  Model $\cup$ Dataset)& $7.18 \times 10^8$ & $8.31\%$ \\
    \bottomrule 
  \end{tabular}
\end{table}
\paragraph{\bf Model and Dataset Density Patterns.}
The bottom of Table~\ref{tab:citation-density-summary}, contains the density of the Model and Dataset relatedness over our table graph.  The Model relatedness is the sparsest, as expected, and the most precise in terms of semantics. The Dataset relatedness is also sparse, but represents a different semantics of relatedness that should be included depending on the task.

\begin{figure}[!htbp]
\centering
\includegraphics[width=1.0\linewidth]{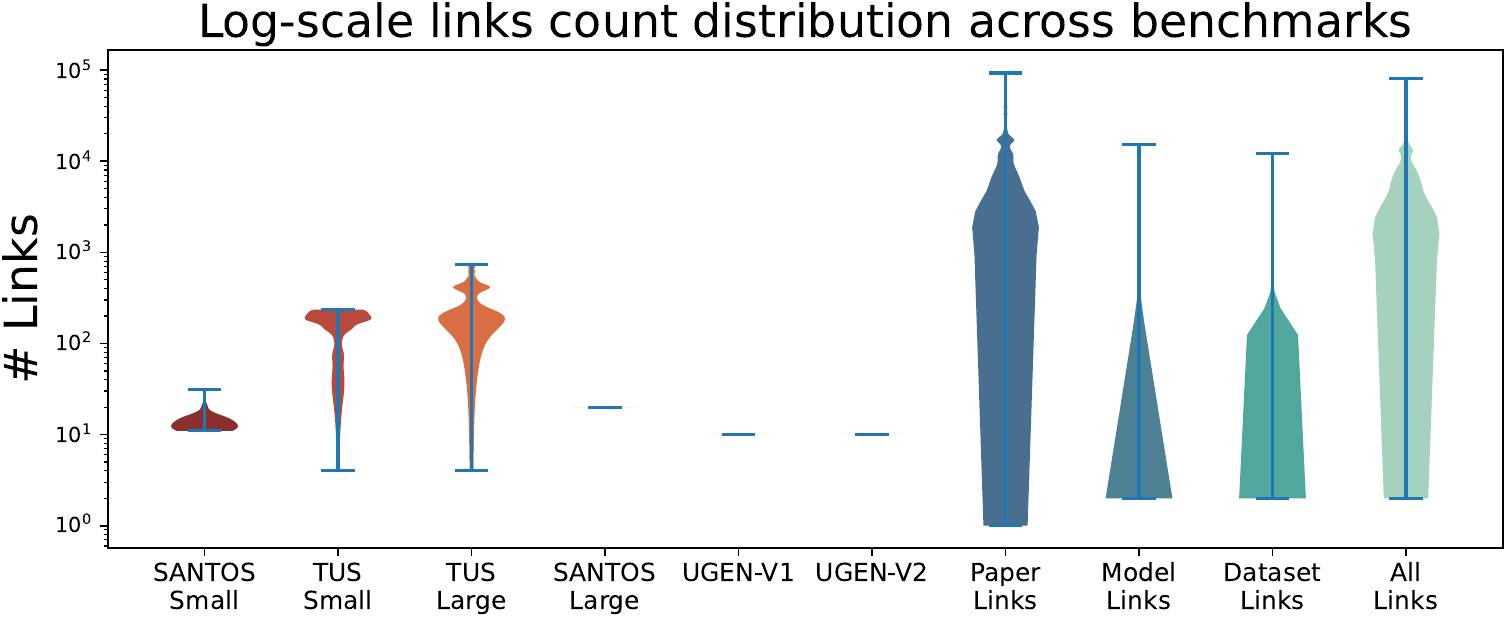}
\caption{
The pronounced long-tail distribution in our ground truth arises from influential models acting as hubs, which link to many model cards.
}
\label{fig:gt_boxplot}
\end{figure}

\paragraph{\bf Table Relatedness Distribution.}
Using the table-as-query paradigm common in data lakes~\cite{DBLP:journals/pvldb/NargesianZMPA19}, 
we further analyze the distribution of related-table counts per query, i.e., how many related tables each query table has (Figure~\ref{fig:gt_boxplot}).
Unlike existing benchmarks such as SANTOS~\cite{2023_khatiwada_santos} and TUS~\cite{nargesian2018table}, which define a fixed set of queries, our benchmark treats every table as a potential query.\footnote{UGEN is excluded because every query has exactly 10 related tables.}
This design yields a substantially larger query space, though it does not imply that the underlying distribution is entirely different.
Indeed, our benchmark exhibits a broader and more variable long-tailed pattern, with related-table counts ranging from only a few to well over a thousand.
This pattern arises from hub effects, highly cited papers or widely reused base models creating dense clusters of connections.
The resulting variability reflects the realistic, heterogeneous nature of the scientific table ecosystem, providing a challenging yet representative ground truth for evaluating retrieval and discovery methods.

%% file: src/experiment.tex
\section{
Evaluation of Table Search Using ModelTables}
\label{sec:experiments}

We now present a detailed experimental result developed with our benchmark.  In this study, we evaluate to what extend existing related-table search methods can find tables that are related using any of our ground truth measures (model, data, or paper).  

\begin{figure}[!htbp]
  \centering
  \includegraphics[width=\columnwidth]{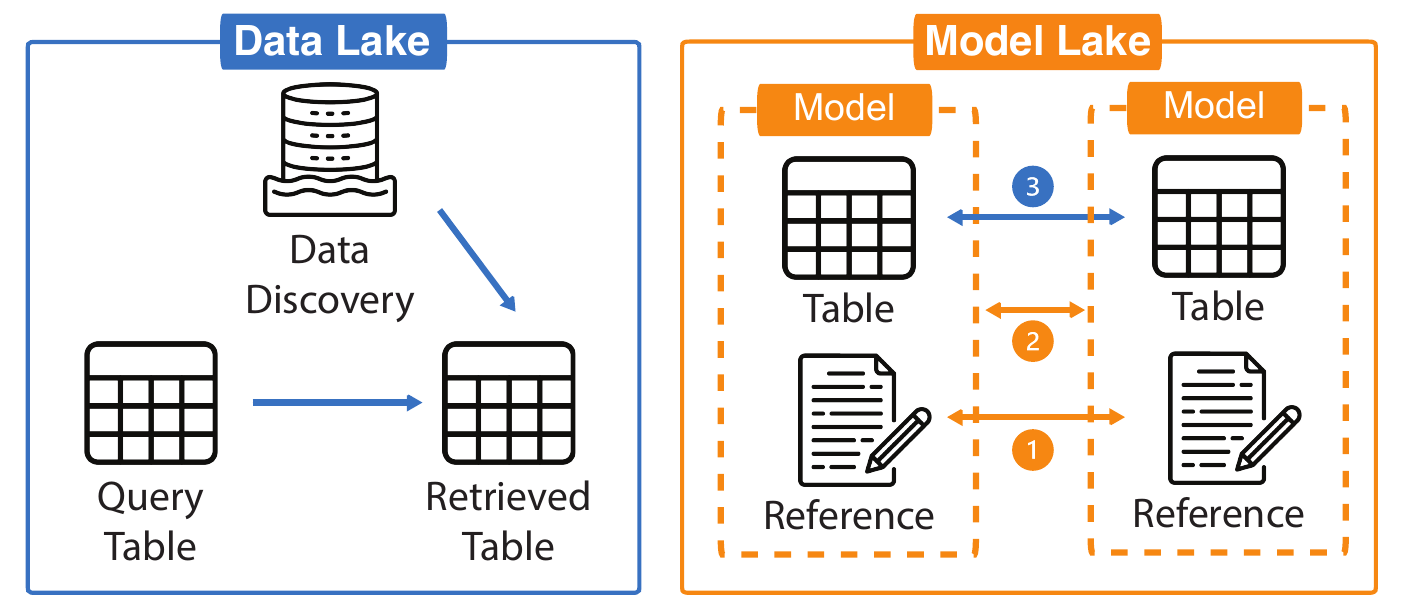}
  \caption{
Grounding the table discovery task in the Model Lake context. To give meaning to a traditional table search (left), we establish a ground truth for relatedness (right) by leveraging the rich connections between associated papers, models, and tables.
}
  \label{fig:idea}
\end{figure}

\begin{table*}[!htbp]
\centering
\small
\caption{Comparative Implementation and Settings for Table Discovery and Retrieval Methods}
\label{tab:combined_method_comparison}
\begin{tabular}{l ccc ccc}
\toprule
\multicolumn{1}{c}{} & \multicolumn{3}{c}{\textbf{Table Discovery Methods}} & \multicolumn{3}{c}{\textbf{Retrieval Methods}} \\
\cmidrule(lr){2-4} \cmidrule(lr){5-7}
\textbf{Method} & \textbf{Keyword Search} & \textbf{Joinable Search} & \textbf{Unionable Search} & \textbf{Dense Retrieval} & \textbf{Sparse Retrieval} & \textbf{Hybrid Retrieval} \\
\midrule
Content & Table & Table & Table & Table & Metadata & Metadata \\
Tool & Blend~\cite{esmailoghli2025blend}  & Blend~\cite{esmailoghli2025blend}& Starmie~\cite{DBLP:journals/pvldb/FanWLZM23}  & FAISS~\cite{douze2024faiss} & Pyserini~\cite{Lin_etal_SIGIR2021_Pyserini} & Pyserini~\cite{Lin_etal_SIGIR2021_Pyserini} + FAISS~\cite{douze2024faiss} \\
Headers & Yes & No & No & Yes & - & - \\
\bottomrule
\end{tabular}
\end{table*}

This section presents a comprehensive evaluation of multiple table discovery methods on our newly constructed ModelTables benchmark. 
We investigate how well existing techniques—ranging from advanced table discovery techniques to traditional embedding-based and metadata-driven retrieval from IR—perform over model tables.

\subsection{%Table Discovery 
Experimental Methods}\label{sec:tableDiscovery}

We evaluate state-of-the-art 
table search and retrieval methods on our benchmark.  We consider data discovery methods including joinable and unionable table search.  
We also consider state-of-the-art table retrieval methods. 
Key implementation characteristics distinguishing these methods are summarized in Table \ref{tab:combined_method_comparison}.

\paragraph{\bf Keyword Search} We use Blend's implementation of keyword search~\cite{esmailoghli2025blend}.  Blend is a recent unified table discovery system with a collection of discovery operators that include keyword search as well as join, correlation, and union discovery.  For Keyword Search, we retrieve tables by matching query table header tokens against candidate table headers and cells.  We consider a table relevant if 
a query token is contained within a candidate's content, with candidates then ranked by hit count.

\paragraph{\bf Join Search} We use Blend's implementation of join  search~\cite{esmailoghli2025blend}.  Given a query table, we search for tables that join on the right-most column (typically a key column in model lake tables).  Following the literature, we do not use headers in this search as the attributes may have different names and still be joinable.

\paragraph{\bf Union Search} We use Starmie's implementation of union search~\cite{nargesian2018table}.  Unlike Blend which only considers exact matching of table values, Starmie considers semantic matching and out-performs Blend on this task. 

\paragraph{\bf Dense Retrieval} 
We create a table embedding that encodes each table (values and headers) as a single vector using a pretrained Sentence-BERT encoder~\cite{reimers2019sentence}. 
Candidate tables are then ranked by the cosine similarity of their embeddings, efficiently performed using FAISS~\cite{douze2024faiss}.

\paragraph{\bf Sparse Retrieval}  Sparse retrieval is a traditional IR technique that relies on term-matching and has been used for table retrieval by applying matching on the metadata describing a table~\cite{yang2018anserini, mackenzie2023efficient}.  
Such metadata retrieval leverages the textual context surrounding each table rather than its internal structure. 
Our preprocessing aggregates raw README text (for Markdown tables) or captions combined with in-text mentions (for \arx\ tables) to form a local "in-context" metadata corpus per table. 
Notably, when a table is associated with multiple metadata texts (due to duplication), all these texts are concatenated to create an augmented corpus for indexing.
Queries are generated by serializing these metadata texts into prompts. 
For sparese retrieval, we use 
Pyserini~\cite{Lin_etal_SIGIR2021_Pyserini} sparse-term index with 1024-clause truncation, and Hybrid Retrieval, which first retrieves the top-100 candidates via sparse search and then applies dense Sentence-BERT re-ranking. 

\paragraph{\bf Hybrid Retrieval} Hybrid retrieval combines the sparse retrieval with dense retrieval based on metadata in a cascaded manner.
Specifically, we first apply sparse metadata retrieval to obtain the top-100 candidates, and then use dense retrieval to select the top-1 result from these candidates.
This dual approach allows us to compare pure sparse retrieval against the benefits of the cascade hybrid approach.

\subsection{%Benchmark 
Evaluation Results}

\paragraph{\bf Setting. } In our evaluation, we use the original ModelTable corpus as the default retrieval candidate pool.
In the augmentation ablation study, we union these table with both the transpose augmentation and the header augmentation (Section~\ref{sec:augmentation}).  
Hence, for each query table in these ablation settings, we  (in parallel) search for the table, its transpose, and its header augmentation.
A retrieval is considered a success if any of these searches yields a top-1 result that corresponds to a correct ground truth table.  In this study, we use \text{Direct} paper relatedness.  
This optimistic evaluation setting is designed to estimate the potential upper bound of performance  achievable by each of the current state-of-the-art methods.

\paragraph{\bf Findings}
\label{sec:main_result}

\begin{table*}[!htbp]
\centering
\caption{Precision@1 of retrieval and table discovery methods across different ground truths (GTs)}
\label{tab:baseline-vs-starmie}
\begin{tabular}{lcccclcccc}
\toprule
\multicolumn{5}{c}{\textbf{Retrieval Methods}} & \multicolumn{5}{c}{\textbf{Table Discovery Methods}} \\
\cmidrule(lr){1-5} \cmidrule(lr){6-10}
\textbf{Method / GT} & \textbf{Paper} & \textbf{Model} & \textbf{Dataset} & \textbf{All}
& \textbf{Method / GT} & \textbf{Paper} & \textbf{Model} & \textbf{Dataset} & \textbf{All} \\
\midrule
Dense Retrieval  & {\bf 0.6651}  & 0.4175  & 0.4143  & {\bf 0.6672}  & Keyword Search  & 0.2043  & 0.0472  & 0.0460  & 0.2060 \\ 
Sparse Retrieval  & 0.5131  & 0.3724  & 0.3749  & 0.5157  & Joinable Search  & 0.2742  & 0.0694  & 0.0723  & 0.2755 \\ 
Hybrid Retrieval  & 0.5410  & {\bf 0.4572}  & {\bf 0.4592}  & 0.5428  & Unionable Search  &  {\em 0.5465}  & 0.3133  & 0.3068  & {\em 0.5487} \\ 
\bottomrule
\end{tabular}
\end{table*}

This section presents the baseline retrieval performance of different methods.
The goal is to establish reference points for future improvements and to understand how different data characteristics impact retrieval effectiveness.

\paragraph {\bf F1: While All Modern Methods Significantly Outperform Keyword Search, 
IR methods Outperform Data Lake Methods except Union Search.}
Our findings (Table~\ref{tab:baseline-vs-starmie}) demonstrate that modern semantic retrieval methods—encompassing approaches based on table content, contextual metadata, and schema-awareness—consistently and substantially outperform traditional keyword search 
across all defined ground truths (GTs). 
Nevertheless, their efficacy is nuanced, varying significantly with the specific semantic signals each method prioritizes.

Dense retrieval on tables proved highly effective within the Retrieval Methods, achieving top performance on the Paper GT and the best result overall, which underscores the importance of capturing the semantics of tables.
In turn, contextual metadata demonstrated its criticality for specific tasks, with Hybrid retrieval excelling on both Model GT and Dataset GT by leveraging essential contextual information beyond raw table data.
While unionable method Starmie demonstrated strong performance, leading the Table Discovery Methods across all GTs.  Notably, Starmie was second best for the Paper and All GT
indicating unionability is a strong signal for \text{Direct} paper relatedness.
However, it was generally surpassed by Dense and Hybrid retrieval when considering overall or other specific GTs, suggesting potential for enhancement by integrating these semantic signals.
And importantly, the tables retrieve by these approaches are different as we show later, further indicating an opportunity for additional research to combine their strengths.

\begin{table}[!hbp]
  \centering
  \caption{Precision@1 of %Starmie 
  table union search across table resources on different GTs.}
  \label{tab:ablation-resource}
  \begin{tabular}{lcccc}
    \toprule
\textbf{Resource / GT}     & \textbf{Paper} & \textbf{Model} & \textbf{Dataset} & \textbf{All} \\
\midrule
ModelCard (M)  & 0.7857  & 0.2599  & 0.2142  & 0.7873 \\ 
GitHub (G)       & 0.8177  & 0.7381  & 0.7366  & 0.8171 \\ 
ArXiv (A)         & 0.4504  & 0.3545  & 0.3607  & 0.4517 \\ 
Semantic Scholar (SS)        & 0.3059  & 0.2630  & 0.2637  & 0.3062 \\ 
\midrule
M+G        & 0.7801  & 0.2766  & 0.2358  & 0.7816 \\ 
M+G+A      & 0.5465  & 0.3133  & 0.3068  & 0.5487 \\ 
M+G+A+SS    & 0.4404  & 0.2700  & 0.2657  & 0.4423 \\ 
    \bottomrule
  \end{tabular}
\end{table}

\begin{table*}[!htbp]
    \centering
    \caption{Precision@1 of methods on augmented tables across different GTs. 
    %\textcolor{red}{ToBeUpdated}
    }
    \label{tab:ablation-aug-baseline}
    %\begin{tabular}{l@{\hspace{1em}}cccc@{\hspace{2em}}cccc}
    \begin{tabular}{l cccc cccc cccc}
    \toprule
    \multicolumn{1}{c}{} & \multicolumn{4}{c}{\textbf{Header-to-Cell Augmentation}} & \multicolumn{4}{c}{\textbf{Transpose Augmentation}} & \multicolumn{4}{c}{\textbf{Both Augmentation}} \\
    \cmidrule(lr){2-5} \cmidrule(lr){6-9} \cmidrule(lr){10-13}
    \textbf{Method / GT} & \textbf{Paper} & \textbf{Model} & \textbf{Dataset} & \textbf{All} & \textbf{Paper} & \textbf{Model} & \textbf{Dataset} & \textbf{All} & \textbf{Paper} & \textbf{Model} & \textbf{Dataset} & \textbf{All} \\
    \midrule
    Dense Retrieval           & 0.7778 & 0.5121 & 0.5042 & 0.7798 & 0.7602 & 0.4950 & 0.4866 & 0.7623 & 0.8142 & 0.5459 & 0.5358 & 0.8160\\
    Keyword Search           & 0.2738 & 0.0512 & 0.0502 & 0.2761 & 0.2989 & 0.0785 & 0.0782 & 0.3012 & 0.4653&0.2365&0.2473&0.4688\\
    Joinable Search & 0.4569 & 0.1541 & 0.1555 & 0.4593 & 0.2751 & 0.0527 & 0.0592 & 0.2776 & 0.5056&0.1617&0.1651&0.5084\\
    \bottomrule
    \end{tabular}
\end{table*}

\paragraph {\bf F2: Significant Performance Variation Arises from Differences in Table Source Quality.}
The source of a table significantly impacts its style and data quality. For instance, tables from GitHub and model cards are often rich with developer-curated metadata, while those from academic papers \arx\ and Semantic Scholar are more structurally diverse. 
To analyze this, we evaluated the performance of Unionable search on each source individually and in combination (Table \ref{tab:ablation-resource}).

Our analysis reveals a dramatic variation in precision based on these sources. 
Model cards and GitHub tables consistently yield high precision, likely benefiting from their well-curated Markdown format. Note that the Semantic Scholar tables consists of LLM-refined plain text rather than structurally rich HTML or Markdown tables.  Hence, the tables are noisier and this dramatically effects the quality of retrieval. 

This principle of the impact of the source characteristics is further confirmed by our combined-source experiments, which underscore that data homogeneity is critical for achieving high precision. 
Performance is strongest for the model cards and GitHub (M+G) subset, where tables share a consistent Markdown format. 
Accuracy progressively declines as less-structured \arx\ tables are introduced (in M+G+A), and further diminishes with the inclusion of the even more varied Semantic Scholar tables (in M+G+A+S).

\paragraph {\bf F3: Domain-Specific Augmentations Enhance the Searchability of Tables.}
To improve retrieval (especially for numerical tables, which often lack explicit semantic context and have layout variations), we evaluated separately our two data augmentation strategies: adding semantic context (Header Augmentation) and normalizing the layout (Transpose Augmentation).
Note that metadata-based retrieval methods are excluded from this analysis, as they do not benefit from table augmentation. Unionable search is excluded, as Starmie's runtime depends on column count, which explodes under the transpose augmentation.
Our analysis (Table \ref{tab:ablation-aug-baseline}) shows the effectiveness of each augmentation.

Adding semantic context proved highly effective
as it
improved precision across all ground truths relatedness (paper, model, and dataset). This confirms that making the implicit meaning of numerical cells explicit is valuable for model lake table retrieval. In  contrast, the Transpose Augmentation consistently improves performance only slightly. 
This demonstrates that the best augmentation strategies for model lake tables  is to enrich the table's existing semantics.

\begin{table}[htbp]
  \centering
  \caption{Precision@1 of unionable search for citation augmentation on Paper GT. }
  \label{tab:ablation-direct-citation-types}
  \begin{tabular}{lcc} 
    \toprule 
    \textbf{Citation Type} & \textbf{Direct} & \textbf{Overlap}\\ 
    \midrule 
All                  & 0.5465 & 0.7487       \\ 
Intent              & 0.5278   & 0.6471    \\ 
Influential          & 0.5097 & 0.5533       \\ 
Intent $\cap$ Influential & 0.5064 & 0.5382\\ 
    \bottomrule 
  \end{tabular}
\end{table}

\paragraph {\bf F4: Robustness and Consistent Performance are Maintained Across Various Ground-Truths.}
Our benchmark permits the use of different citation semantics in establishing paper relatedness (Table~\ref{tab:citation-density-summary}).  Table \ref{tab:ablation-direct-citation-types} presents the precision of union search over eight different citation definitions.  
Precision predictably drops when stricter GT criteria are applied (meaning fewer citation relationships), a direct consequence of the ground truth becoming sparser. 
This inverse relationship is particularly evident with Paper GT, where precision drops as citation criteria become more stringent.
Interestingly, when comparing GTs, those defined by overlap consistently yield higher precision than GTs based on direct citation criteria, suggesting overlap inherently captures a broader, denser set of relevant connections.

These findings underscore that while retrieval performance metrics are highly sensitive to the definition and statistical properties of the ground truth itself, the methods exhibit consistent behavior even as ground truth criteria become more stringent. 
This highlights the importance of understanding how GT selection influences reported performance. 

\paragraph {\bf F5: Structural augmentations are effective in improving retrieval robustness alongside semantic augmentations.}

In addition to semantic and citation augmentations, we investigate the impact of table structure perturbations, including column and row shuffling. These augmentations are evaluated using  Starmie only, as its encoder is specifically designed for structural robustness. Column shuffling at inference yields the largest gains, highlighting that augmenting table structure is a practical strategy for robust retrieval across heterogeneous model lakes.

We observe that all structural augmentation tricks improve retrieval performance, confirming substantial variation in table styles across authors and sources. 
Column shuffling provides the largest gains, highlighting that column order is especially arbitrary—particularly in academic tables with semi-structured formats, where unionability is difficult to detect. 
Structural augmentations, especially at the column level, help the encoder focus on semantic and schema similarities rather than layout, resulting in more robust retrieval in heterogeneous model lakes.

\begin{table}[htbp]
  \centering
  \caption{Precision@1 of Starmie with structural augmentations across different GTs.}
  \label{tab:ablation-augment}
  \begin{tabular}{lcccc}
    \toprule
    \textbf{Augmentation / GT}     & \textbf{Paper}  & \textbf{Model} & \textbf{Dataset} & \textbf{All} \\
    \midrule
    Base        & 0.4119& 0.1899  & 0.1806& 0.4133 \\
    Drop Cell  &0.4607& 0.2323 &0.2240& 0.4627\\
    Shuffle Col & \textbf{0.5465} & \textbf{0.3133} & \textbf{0.3068}& \textbf{0.5487} \\
    Shuffle Row     & 0.4419 &0.2031  &  0.1979&0.4436\\
    \bottomrule
  \end{tabular}
\end{table}

\subsection{Example Studies}
None of the known retrieval methods has great performance on ModelTables even for the simple precision\@1 task.  To shed better light on why, we give two examples of tables retrieved using a structural technique (unionable search) and tables retrieved using table-embeddings (dense retrieval).  We use these examples to motivate intuitively why we believe our study motivates the need for careful combination of structural and semantic signals to improve related table search.  

\paragraph{\bf Example I: Unionable Search Example}

\begin{figure*}[!htbp]
  \centering
  \includegraphics[width=\linewidth]{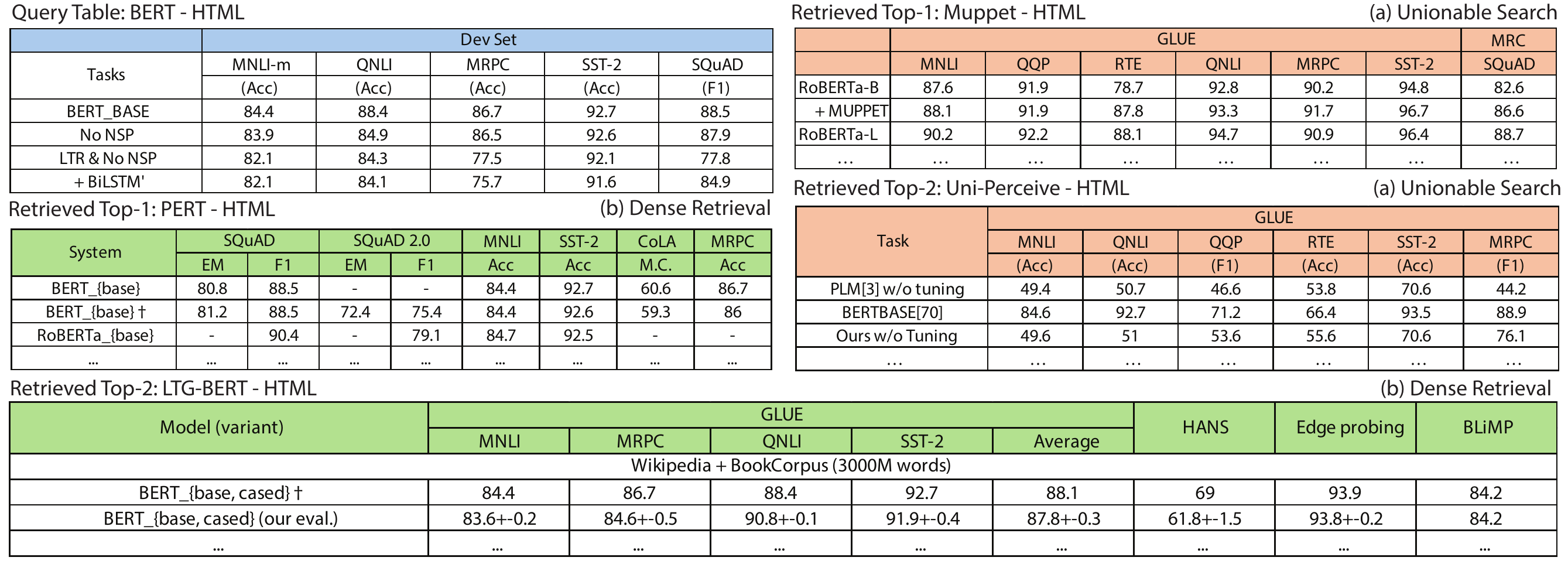}
  \caption{Example of table discovery using (a) union-based and (b) dense retrieval for a BERT query.
(a) Union-based retrieval returns schema-aligned tables with consistent reporting of GLUE~\cite{wang2018glue} and SQuAD~\cite{rajpurkar2016squad} results.
(b) Dense retrieval yields more diverse and semi-structured tables, reflecting varied author practices in reporting and layout.
  }
  \label{fig:result_example_all}
\end{figure*}

Here we demonstrate using unionable search for GLUE-style performance tables, showcasing the search’s ability to retrieve tables that 
describe related models.
Figure \ref{fig:result_example_all} (a) illustrates a BERT query table and the Top-2 retrieved tables using Starmie (union table search.   these search results are clearly relevant and enable direct, side-by-side comparison of both baseline and advanced models (e.g., BERT~\cite{devlin2019bert}, RoBERTa~\cite{liu2019roberta}, MUPPET~\cite{aghajanyan2021muppet}, Uni-Perceive~\cite{zhu2022uni}) across a consistent set of NLP tasks and evaluation metrics such as MNLI, QNLI, MRPC, SST-2, and SQuAD~\cite{rajpurkar2016squad}.  
These tables exhibit clear structural and semantic alignment, reporting results on overlapping benchmarks that make them unionable.  But they are also clearly relevant to a data scientist analyzing model behavior.

Our observations highlight the limitations of traditional text-based information retrieval methods, which can only address coarse-grained questions such as “Which models are available for a given task?”, falling short on fine-grained performance comparisons at the metric level. 
Using table search,
we can retrieve related and integratable tables that expand the analysis that can be done with the query table alone.
This approach enables users to find tables that help them to make precise, context-aware decisions grounded in detailed performance data from multiple models and benchmarks.

\paragraph{\bf Example II: Dense Retrieval Example}

As shown in Figure \ref{fig:result_example_all}(b), dense retrieval uncovers tables that, while still centered on BERT variants and key NLP benchmarks such as GLUE and SQuAD, exhibit much greater heterogeneity in schema and reporting style compared to union-based results. Unlike the more schema-aligned tables surfaced by union search, dense retrieval often returns tables with alternative column organizations, extra metrics, or less conventional formatting—reflecting the diverse ways authors present experimental results.

This flexibility can be seen as a major advantage: dense retrieval is able to connect tables that are semantically related but do not share  structure or use differing conventions. 
As a result, it can successfully retrieve relevant data even when column names, task sets, or layouts diverge from the query. 
This capability is especially valuable for scientific corpora where reporting standards vary widely and metadata may be noisy or incomplete.

However, dense retrieval is not without limitations. Because it relies on semantic similarity, it tends to retrieve tables with highly overlapping baselines or benchmarks—most notably, results focused on "BERT-base"—rather than a broader variety of model variants or experimental settings. This may result in a concentration of retrieved tables that report on the same baselines, potentially missing tables that, while structurally unionable, cover a wider range of models or alternative approaches. In contrast, union-based methods, are more likely to surface diverse model variants and experimental configurations, supporting broader comparison across methods.

In summary, dense retrieval effectively complements union-based methods by bridging gaps in schema alignment and surfacing structurally diverse, semantically relevant tables. 
However, it is less effective in promoting methodological or model diversity, often prioritizing tables with significant overlap in baselines and benchmarks. 
This suggests there are open research challenges in combining the strengths of both approaches to achieve better model lake performance.

%% file: src/applications.tex
\section{Applications}\label{sec:apps}

Beyond related table search, our benchmark has numerous applications.  We elaborate on just a few here.

\noindent{\bf Model Search}
Unlike previous table corpora, ModelTables can be used for model understanding.  One possible line of inquiry is to see if table relatedness can be used as a proxy for model relatedness.  Currently, model lakes rely on keyword search over names and metadata.  Semantic search is only just emerging and relies on LLM Q\&A over open models.
We hypothesize that tables describing related models exhibit high structural and semantic similarity, allowing table discovery to serve as a proxy for identifying related models.
Scientific tables have been shown to encode rich semantics useful for reasoning and retrieval~\cite{kruit2020tab2know,wu2023column,zhang2024scitat,lu2023scitab,desai2021tablex}.
Validating this hypothesis requires (1) robust table-retrieval methods and (2) a principled ground-truth definition of model relatedness.

\noindent{\bf Model Table Integration}
While the problem of integrating a set of data lake tables has been studied~\cite{khatiwada2022integrating}, the problem of integrating inconsistent data lake tables, including tables with many numeric values has not received sufficient attention.  Using ModelTables, we can identify a set of tables about a model.  We can characterize them into configuration tables, performance tables, and other types of tables.  Within each type of table, we can identify inconsistencies in the tables and use integration techniques to resolve inconsistencies and define a best integration.  We speculate that such research could have implications not only for data integration research, but for research on model card verification and generation~\cite{DBLP:conf/naacl/0004LJD24}.

\noindent{\bf Model Understanding}
One of the holy grails in model lake research is understanding which model best suits a task~\cite{DBLP:conf/edbt/PalBM25}.  To this end, we speculate that using principled search over Model Tables, we may be able to understand which model performs best on different benchmarks and synthesize this rich structured information to do better query answering over models.    

%% file: src/conclusion.tex
\section{Conclusion \& Future Work}\label{sec:conclusion}

\begin{figure}[!htbp]
  \centering
  \includegraphics[width=\linewidth]{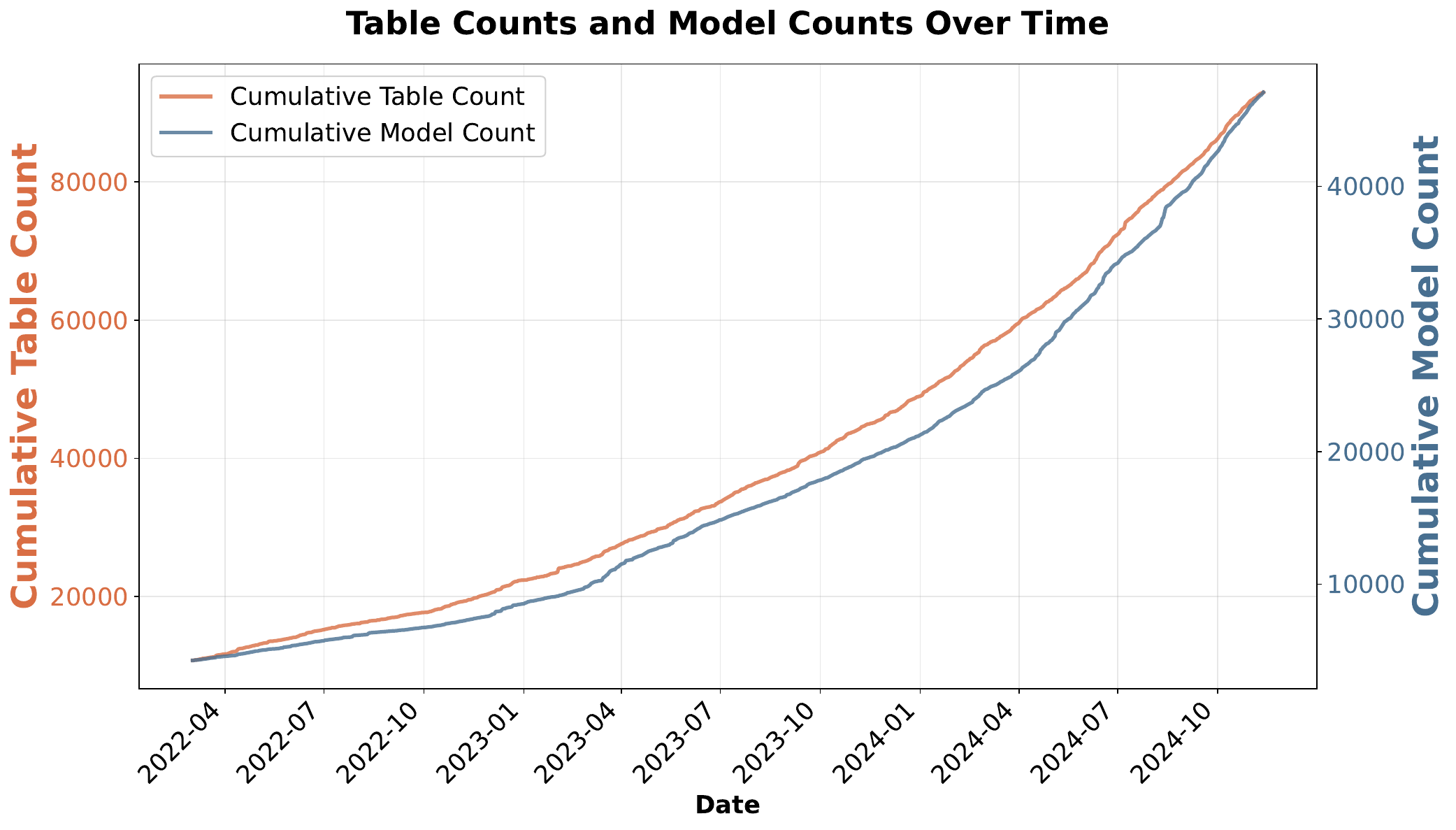}
  \caption{Cumulative counts of Huggingface models and tables 
  over time. 
  }
  \label{fig:table_model_counts_over_time}
\end{figure}

We presented a new benchmarks, ModelTables.  We presented robust and repeatable methods to generate our benchmark and ground truth.  The benchmark itself was created with a January 2025 crawl of Huggingface and related resources.   As shown in Figure~\ref{fig:table_model_counts_over_time}, the number of tables and models are growing exponentially over time, something that will allow the  generation of a significantly larger ModelTables V2 using our reproducible and open benchmark creation pipeline and code.
We presented 
 characteristics of the benchmark tables, and showed that the ground-truth, derived from real reliable signals created by model developers, shows realistic characteristics (e.g., hub popular models and a skewed relatedness distribution) that is unique among all table benchmarks surveyed (which use either human-labels or LLM generated labels).
 We presented one extensive use case of the benchmark to evaluate related table search methods on model tables where unlike in web tables or data lakes, relatedness cannot be defined solely based on structure~\cite{DBLP:conf/sigmod/SarmaFGHLWXY12}.  We showed how our benchmark can give insight into research gaps and new areas for innovation.  We also speculate on several other applications in both table understanding and model understanding.  

